# Morphology of Banded Anisotropic Structures

**Eduardo Leiva M[1] Marco Reyes Huesca[2] and Enrique Geffroy[3]**
[1]Instituto de Investigaciones en Materiales, Universidad Nacional Autónoma de México, Avenida Universidad 3000, Mexico, D.F., 04510, Mexico
[2]Departamento de Termofuidos, Facultad de Ingeniería, Universidad Nacional Autónoma de México, Avenida Universidad 3000, Mexico, D.F., 04510, Mexico

e-mail address: eleiva@iim.unam.mx

The study of *the formation of bands of drops along the direction of the vorticity* in a concentrated emulsion is presented. That is, the observed anisotropic structures are characterized by regions of high concentration of drops interspaced with regions of about equal width where hardly any particles are present. These particle distributions are a clear departure to what has previously reported for this emulsion. These patterns are induced by the effect of a simple flow in circular parallel plate geometry (CSS450-Linkam), but only after a prolonged time after the inception of the flow. That is, the observation of bands in emulsions only appears after various other phenomena occur in the emulsion.

That is, the separation of the flow cell plates is 100 μm, and the observation of bands occurs within the interval of shear rates of 3.0 to 5 s$^{-1}$, but only after a long evolution of the initial structure with a series of flows of 0.75 to 2.25 s$^{-1}$.

**Keywords:** Banded structure, anisotropy, emulsion.

## I. INTRODUCTION

The formation of oriented structures in two-phase systems, as in this study, and in particular the formation of bands of droplets, is a phenomenon that attracts considerable interest in the scientific and industrial fields. Its vast set of applications is directed to the fields of microfluidics, the food industry, as well as granular materials, among other possibilities.

Some band formation studies have been reported in the literature for different types of samples, such as granular materials [1], stiff particle suspensions [2], liquid crystals , polymer solutions [3] and attractive emulsions [4] [5]. The formation of bands in dilute emulsions is reported mainly in the geometry of concentric rollers [6], where band formation can occur in all three directions (the direction the flow, of the velocity gradient, and in the direction of vorticity). These bands may depend on the flow of different variables and under particular and different conditions.

The essential difficulty of understanding of band formation in biphasic systems is the simultaneous presence of physicochemical, hydrodynamic, and mechanical effects. Also, It is quite common that banded structure materials show length scales covering several decades. For example, the characteristic length scales observed in composites for carfenders is as small as 0.3 μm while its macroscopic features are more extensive than several millimeters. In technologies of the order of micro and nano, where biphasic mixtures play an essential role, these physical processes are still poorly understood [7].

The trend of studies in this field have been recorded for a couple of decades, and it is considered that it occurs in complex fluids whose relaxation times are slow [8] [9] [10]. However, there are recent reports in which the phenomenon can also be evidenced in relatively "simple" systems, as are two Newtonian liquids, immiscible and without surfactants [11]. Thus, the slow relaxation mechanisms can not be inferred from the dynamics of individual constituents. The slow mechanisms are the result of collective dynamics that are not easily determined or understood. Next,we systematically study the formation of bands in a water-in-oil emulsion, with a 50/50 fraction and under a simple shearing flow.

## II. SHEAR BANDING IN THE EMULSION

The first reports of bands formation by flows emphasize the correlation with the shearing rate in solid samples, then in mixtures of molten polymers and finally in emulsions. However, in the field of emulsions, the work of Caserta, et al. [12] is the most recent and addressing the relationship between structure and flow properties. Consequently, the phenomenon of bands in emulsions is a rather new and poorly studied phenomenon, even though it is highly relevant in advance manufacture of medical and optical devices, whose motivation lies in minimizing its components. From a purely conceptual perspective, the essential objective is to understand the relationship of physical phenomena of large scales with the microscale of its structure, and, more importantly, of the microscale at the macro scale [13]. That is, understanding a coupling of time- and length-scales induced by a non-equilibrium flow condition. First, a brief explanation of the states by which the emulsion must evolve to reach the formation of bands. It is essential to say that the following description is the result of a systematic consultation of observed phenomena related to the morphology of biphasic systems.

When an emulsion is subjected to a pure flow, the observed structural phenomena depend on a set of multiple variables. But to favor the phenomenological understanding of the banding structure, we will focus on three variables specifically; the confinement ($Co = 2R/H$), the viscosity ratio ($p$) and the shear rate ($\dot{\gamma}$). The confinement ($Co$) is the ratio between the diameter of the drop and the separation of the circular plates. This dominant dimensionless scale has its origin on the definition frequently used with diluted systems, for example, an isolated drop. In the literature, we find banded systems with $p \sim 1$ and values of $0.2 < Co < 0.56$ [7]. From which the phenomenological idea is extracted to create the scenarios in the formation of bands.

The scenarios of the morphological evolution induced by the flow are, initially, the alignment and elongation of drops; the second may be the coalescence of drops and the third when highly elongated its rupture into multiple smaller drops. These phenomena may happen simultaneously. Moreover, after a long time, the evolution may show fourth scenario the formation of bands.

The bands appear from an initial emulsion, visibly homogeneous. Prior to the observation of bands, other measurable phenomena dominate the emulsion structural evolution. Bands are ordered, equally spaced, and intercalary along the direction of vorticity. The first visible band is from the outside in, on the vorticity axis. This scenario is for $p = 0.27$ and a critical capillary $Ca_{cr} \sim 0.21 \pm 0.07$, explained next.

### III. THE BREAKUP OF DROPLETS UNDER A SHEARING FLOW

The capillary number for the rupture of a droplet is called the critical capillary, $Ca_{cr}$, It is precisely when the internal stresses of the drop are overcome by the stresses of the flow and break up occurs. In the work of Grace [14] and that of De Bruijn [15], the intrinsic relationship between $Ca_{cr}$ and $p$. Furthermore, Bentley and Leal [16] report the results for droplet deformation and rupture in two-dimensional flows, ranging from single-cut flow to pure elongational flow. Furthermore, for emulsions subjected to simple shear flow flows, Jansen [17] has recently shown that the critical capillary number decreases as well with increments of *the fraction of the disperse phase*; that is: $Ca_{cr}(p, \phi)$.

Single droplet breaking mechanisms and shapes of Newtonian liquid have been extensively studied. If $Ca \ll 1$, the shape of the drop is slightly ellipsoidal, depends on the value of $p$, and the drop is aligned at an angle of 45° with respect to the flow direction. The increase in capillary number is proportional to the elongation of the drop in a stable state, the drop also rotates, aligning along the flow direction, to maintain the stable state. At the moment where the value of the capillary number increases and exceeds the critical value, the rupture of the drop is observed. The different modes of rupture of a droplet depend on the viscosity ratio. For a viscosity ratio of less than one $p < 1$, the droplet acquires an elongated shape with a pointed end; As the tip streaming phenomenon occurs. For $p$ approximately equal to 1, the central portion of the droplet forms a neck (or necks) until followed by the breaking up into two daughter-droplets, with small satellite droplets between them. $Ca \gg Ca_{cr}$, Droplets are deformed into long, thin fiber filaments that eventually break up through the capillary wave instability mechanism. These mechanisms become more complicated as the density of disperse phase drops increases, and length scales of a different object overlap considerably, including non-negligible effects due to the presence of the flow cell walls.

## IV. MATERIALS AND METHODS

### A. Constituents and preparation of the emulsion

A water-in-oil type emulsion is studied; it is composed of a dispersed phase and a continuous phase, aqueous and oily, respectively. The two phases are immiscible and with very similar densities; $\rho_w/\rho_o = 1.03$. The alkane components of the continuous phase are to increase the viscosity of the emulsion.

The dispersed phase has a dynamic viscosity of $0.57\ Pa \cdot s$, and the continuous phase has a dynamic viscosity of $2.08\ Pa \cdot s$ at 30 °C, and a shear rate range of 0.01 to 10 $(s^{-1})$. The densities of the phases are 0.98 $g/cm^3$ and 0.98 $g/cm^3$ at 30 ° C for the dispersed phase and the continuous phase, respectively. These two characteristics make this sample a good working fluid mixture for simulations of oil-water emulsions such as crude oil and water.

The interfacial tension $\sigma$ was determined by the method of deformed drop retraction, DDR, as described by Guido and Villone [18]. The average value of the interfacial tension for 11 droplets is 0.11 mN / m.

Even though a steady change of the drop size distributions occurs previously to the application of a flow with a shear rate of $3.0\ \mathrm{s^{-1}}$, here reported are banded structures after other notable changes are observed, the former having been reported [19]. All flow experiments were performed at a pre-shear rate of the emulsion, which corresponds to a series of steady flows, beginning with a shear rate of $0.75\ \mathrm{s^{-1}}$ up to $3.0\ \mathrm{s^{-1}}$ for $\sim 500$ s each step. Preconditioning flow manifests a complex dynamic for the structure. However, it does not allow inferring a drastic change, such as the subsequent evolution into bands. The loading of the flow cell is carried out as in previously described experiments. After loading, the emulsion was allowed to relax over a period of $\sim 600$ s.

### B. Parallel disks cell/device

All experiments were carried out using the parallel plate geometry (Linkam CSS450, Linkam Scientific Instruments Manufacturer, Tadworth, UK), schematically shown in Figure 1. The upper disk remains still, meanwhile, the lower disk moves to impose a simple shear rate on the emulsion. Observations were conducted at 7.5 mm from the center of rotation of the lower disc. The observation window is 2.8 mm in diameter, and images are captured in the vorticity-velocity observation plane. All measurements reported in this work were made with a 0.1 mm gap between the discs and a temperature of 30 °C.

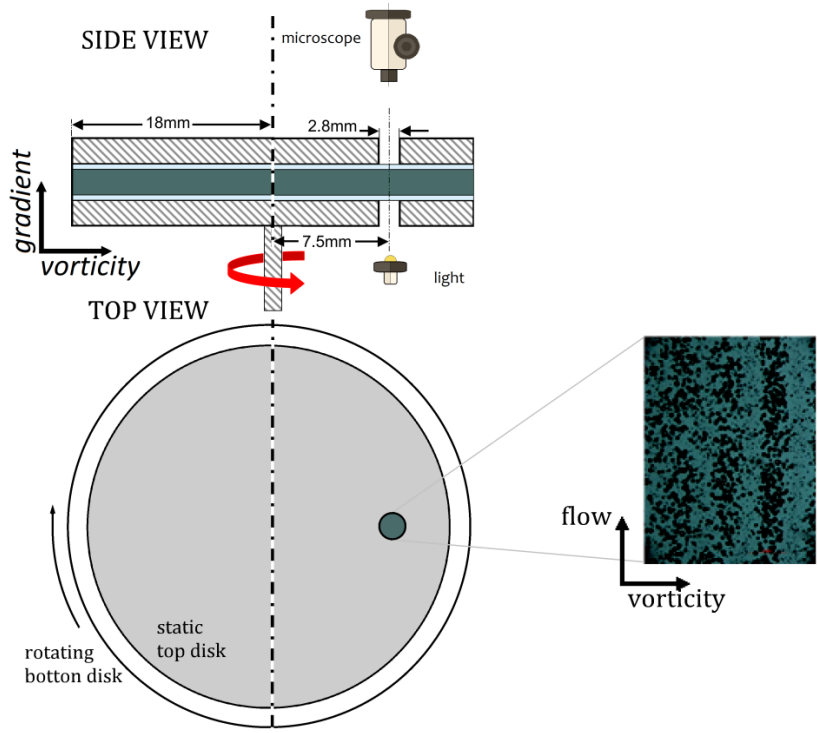

**Figure 1.** Schematic representation of the shearing device: parallel-plate geometry with a diameter of 36 mm and a gap of 0.1 mm. The velocity–vorticity axes describe the observation plane. The lower disc motion imposes a simple shearing field on the sample.

Images of the banded structure of the emulsion are experimentally observed through a small hole of the flow device —which covers a surface of only 0.6% of the total area of the geometry— and captured images (1000 x 1340 μm) corresponds to 0.02 % of the flow field area. Thus, the real extension in the flow of these bands cannot be inferred from these experiments, precluding a better understanding of the role that substantial time- and length-scales may play in this phenomenon. In Appendix C, the location of the captured image in the observation hole is described. The formation of the vorticity axis bands (parallel to the flow axis), as shown in Figure 1, appear to arise in the centripetal direction. The width of the first band changes with time and is related to (is of the order of) the separation of the plates.

Visualization of the microstructure was performed with a Nikon SMZ-U light microscope (manufacturer Nikon Corp., Tokyo, Japan). For image capture, a Nikon Digital

Sight DS-2mV camera was arranged in a bright field lighting arrangement. Images are processed with ImageJ® software (US National Institutes of Health, Bethesda, MD, USA, City, Country), automatically and manually. A full view image was ensured with the focus plane centered on the flow field plane; the image capturing process was optimized to reduce the emulsion turbidity and proper exposure to light [20].

### C. Images analysis

The evolution of the microstructure was visualized with the acquisition of multiple images; in the beginning, at the end of each shear period and when the flow stopped completely. The images were spaced at intervals of 1 s for statistical analysis, as shown in Figure. 2.

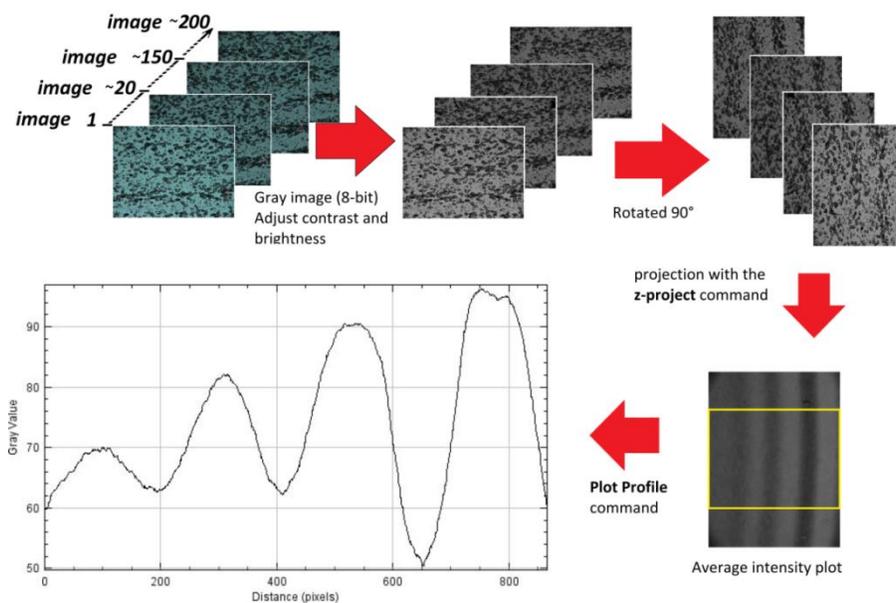

**Figure 2.** Scheme of the image analysis procedure for the description of the evolution of band formation. It is essential to include the flow direction with an arrow. The stacking of images correspond to the shear rate of $5\ s^{-1}$.

To extract the information from the captured images, we proceed in steps. These steps are: (a) The capture: this stage is the experimental arrangement, the camera, the zoom; the details are described in Section 2.6. (b) The pre-processing: where the noise of images is reduced, and those details that are not of interest in the experiment are eliminated. (c) The segmentation: here, the drops to be measured are evaluated according to a size and shape criteria. The extraction of the characteristics: here, we proceed to obtain the chosen measurements of each of the objects, images, and stacks.

The observation of bands in the direction of vorticity is analyzed quantitatively using ImageJ® software and compared systematically against changes due to increments in the applied shear rate. The stage of pre-processing, segmentation, and extraction of the images are shown in Figure 2. The ~200 images captured during the flow are collected for each shear, i.e., six collections of approximately 200 images each. These ~200 images are stacked and rotated 90° in an anti-clockwise direction, to favor the analysis and subsequent visualization of the results. Subsequently, the stack is converted to 8-bit grayscale, also adjusting the contrast and brightness of the stack.

Later, a projection with the z-project command is made, —remember that the z-project command analyzes stacks by applying six different methods of projection of pixels of the stack. The six techniques of the z-project command are shown in Figure 3, where each of the images is the result of applying the z-project command to the same shear rate (of $5 \text{ s}^{-1}$) image. In Figure 3, all six techniques are presented. Each technique allows us to visualize the formation of bands in a different way; for example, in some cases, the band is shown in black and others in white or is not visible.

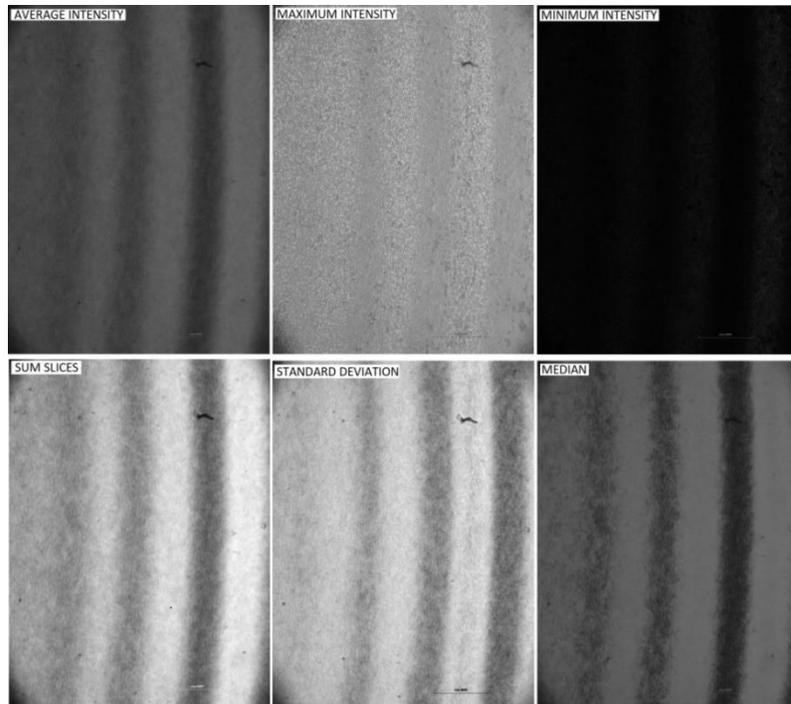

**Figure 3.** The z-project techniques for banding analysis. From left to right: average intensity, maximum intensity, minimum intensity, sum slices, standard deviation, and median.

To choose the most favorable technique for our purposes. Each technique was graphed of intensity vs. radial distance of the whole image is made with the plot profile function, as can be seen in Figure 4. The graph obtained of intensity vs. radial distance shows undulations, where intervals of intensity—where 0 is equal to black and 255 to white— are used.

For example, when using Average, minimum sum and median graphs, valleys represent areas (regions) with the highest droplet population during flow, and crests, vice versa. Conversely, the same happens with the techniques of maximum intensity and standard deviation.

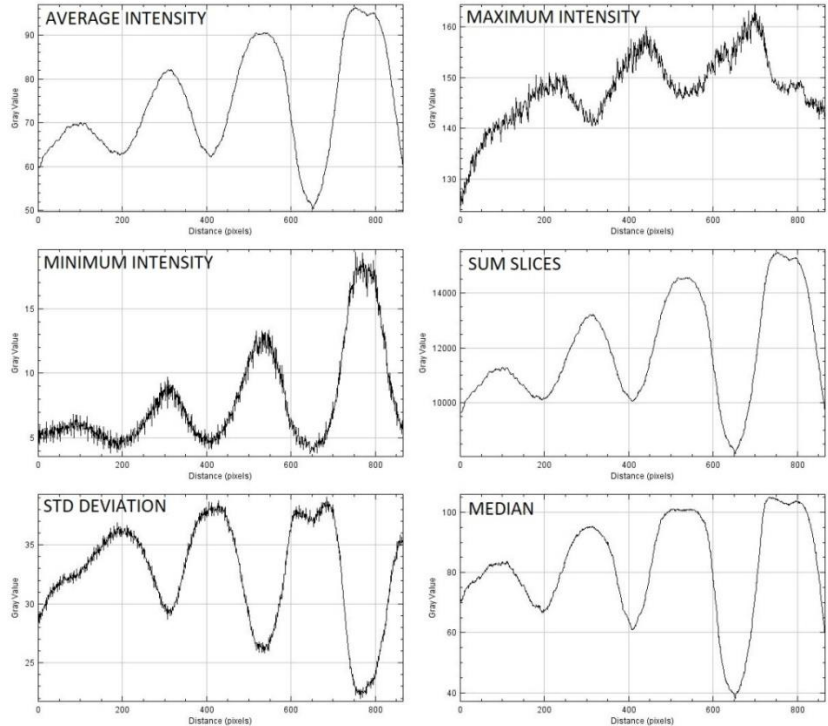

**Figure 4.** The plot of the intensity vs. radial distance graph of z-project techniques for banding analysis. From left to right: average intensity, maximum intensity, minimum intensity, sum slices, standard deviation, and median.

For example, when using Average, minimum sum and median graphs, valleys represent areas (regions) with the highest droplet population during flow, and crests, vice versa. Conversely, the same happens with the techniques of maximum intensity and standard deviation.

Thus, the average intensity technique is chosen here for the analysis, for it provides be the most accurate method for our purposes. The noise is minimal and allows identifying the evolution of the population of droplets at the borders of the bands. To facilitate understanding of the data, the gray hue of the average intensity technique is inverted.

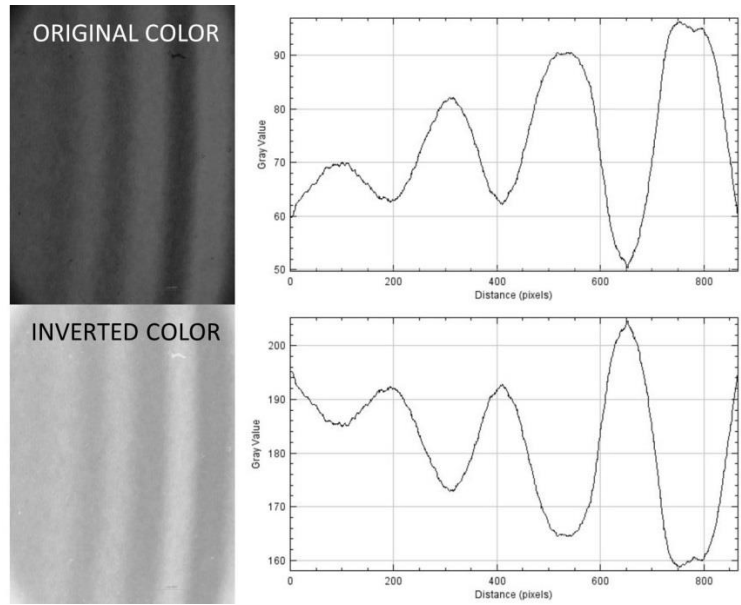

**Figure 5.** Image obtained with the average intensity technique with color inversion (left side) and its graph of intensity vs. radial distance (px) for a shear rate of 25 s$^{-1}$ (right plots). In the image of inverted colors, valleys correspond to a low population of drops and peaks to a high population of drops.

In this manner, at the moment of making intensity vs. radial distance graphs, each valley represents the low population of drops, and each peak the high population of drops, as shown in Figure. 5. Hence, the projection on the perpendicular axis of the image plane is made with the average intensity technique, for each single shear rate.

The described methodology is applied to each of the studied shear rates, as shown in Figure 6.; the abscissa corresponds to the position along the vorticity axis, while the ordinate contains data of the normalized intensities [0, 1]. Image processing provides information on the average position of the peak of the band at all times of the evolution. As well, it is possible to make a qualitative comparison of the number of drops in each zone by evaluating the *relative* height of each peak (wrt. the intensity of the valley). Figure 6 shows the profiles obtained from stacking images for various shear rates. Only shear rates higher than 3.0 s$^{-1}$ the show, clearly, band formation.

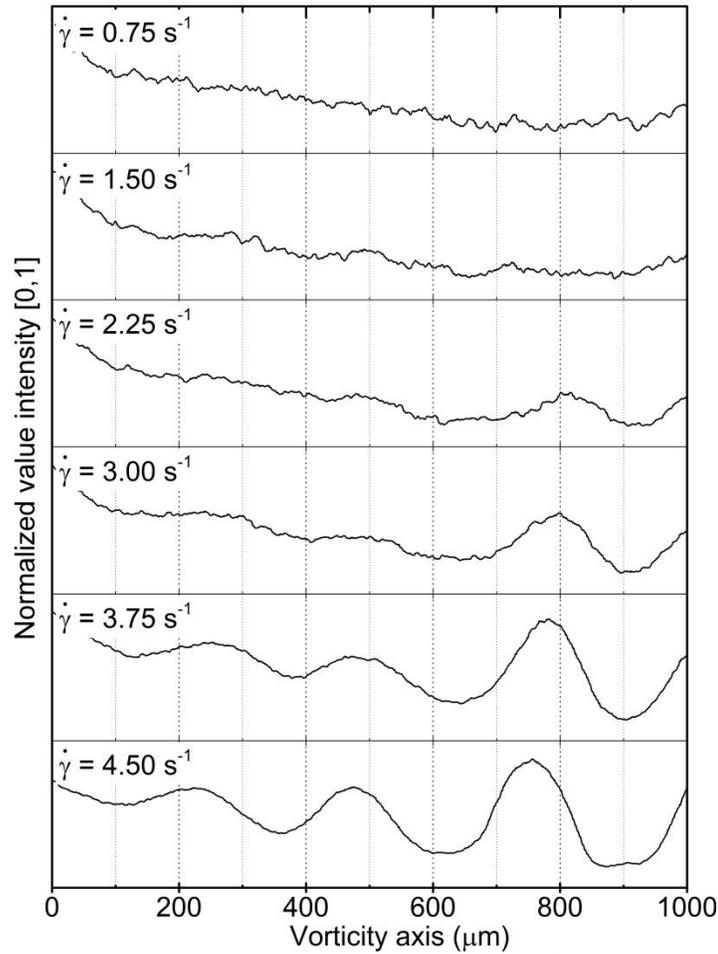

**Figure 6.** Intensity vs. radial distance graph from 0.75 to 5 s$^{-1}$, each 0.75 (top to bottom). Formation and displacement of the bands. The intensity axis was normalized from 0 to 1.

In order to determine the long times evolution of each band, a second statistical analysis is carried out. The objective is to characterize the intensity data by adjusting a Gaussian curve with parameters shown in Figure 7. The amplitude (A), the Full Width at Half Maximum (FWHM), and the position of the peak ($X_c$) are the parameters taken into account for the acquisition of the width of the band, the population of the drops, and the possible displacement of the peak.

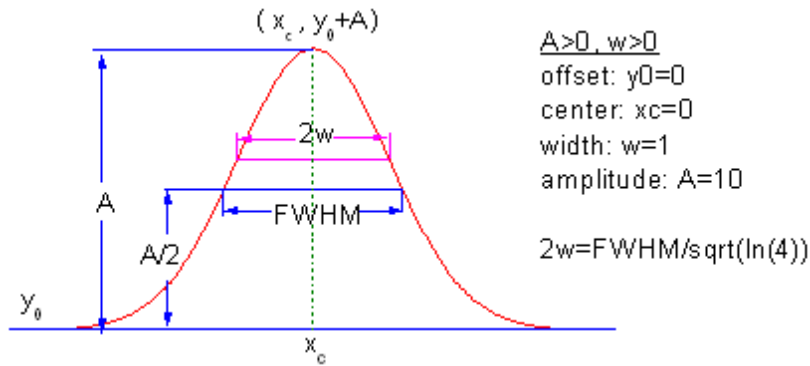

**Figure 7.** Plot the average intensity of 10 images (1 fps) for a shear rate of 5 s$^{-1}$. The parameters obtained from the adjustment of a Gaussian curve are indicated.

Figure 8 shows the four parameters that characterized the dynamics of each peak. Each intensity vs. position plot is composed of 867 columns (aligned with the flow direction), and each column has 1155 intensity values. Thus, each plot of the graph —plots Figure 8b and 8d— represents the mean value of the intensity for 11,550 readings produced with ten images, at a specific time in the evolution of the drops distribution.

The black trace in Figure 8a corresponds to the mean intensity value for that column. Intense (and light) red bands show the uncertainty range at each column, indicating that large fluctuations are due to a large number of drops passing by. However, as is shown in Figure 6, fitting a Gaussian curve to peaks on intensity plots requires setting a common baseline at both ends of the curve, i.e., traces shown in Figure 8c and 8e.

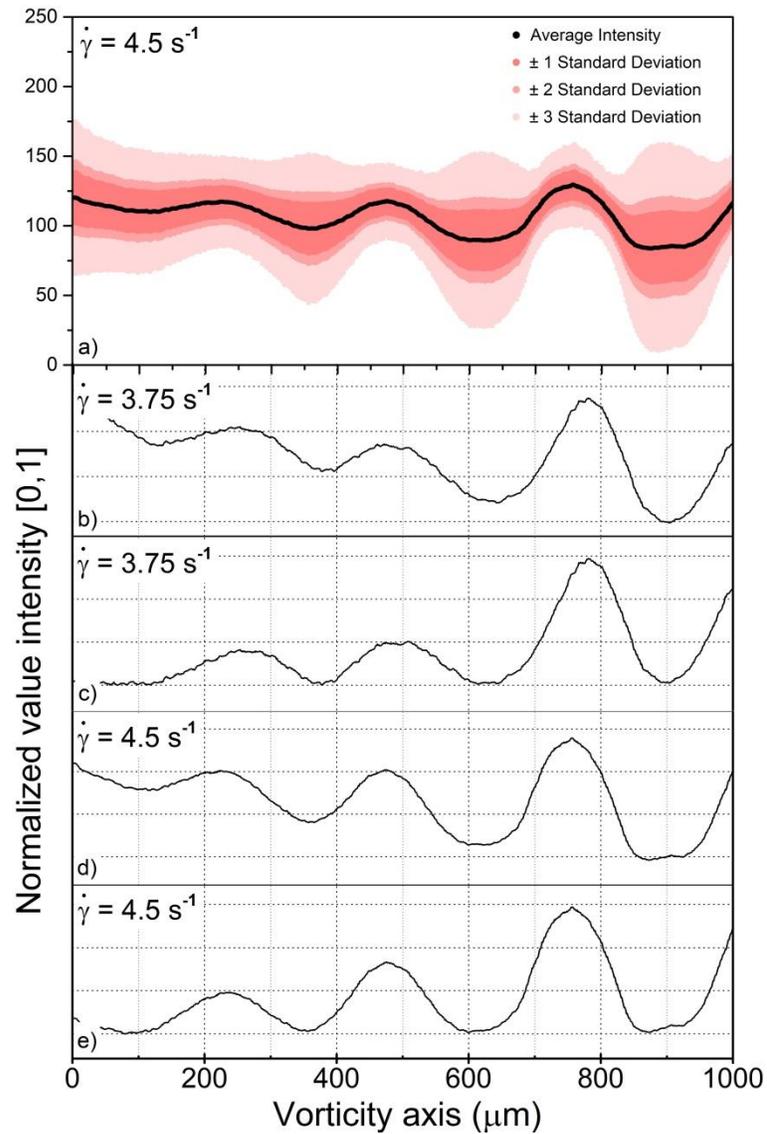

**Figure 8.** Intensity vs. vorticity axis (μm) of a pack of 10 images (1 fps) for $\dot{\gamma} = 3.0 \text{ s}^{-1}$ and $5 \text{ s}^{-1}$. Plot (a) shows the black trace for the mean intensity at each column, while red bands correspond to their uncertainties. Plots (b) and (d) correspond to raw intensities, while (c) and (e) to intensities with a baseline correction.

Thus, Figure 9 shows the adjustment carried out for the two highest shear rates.

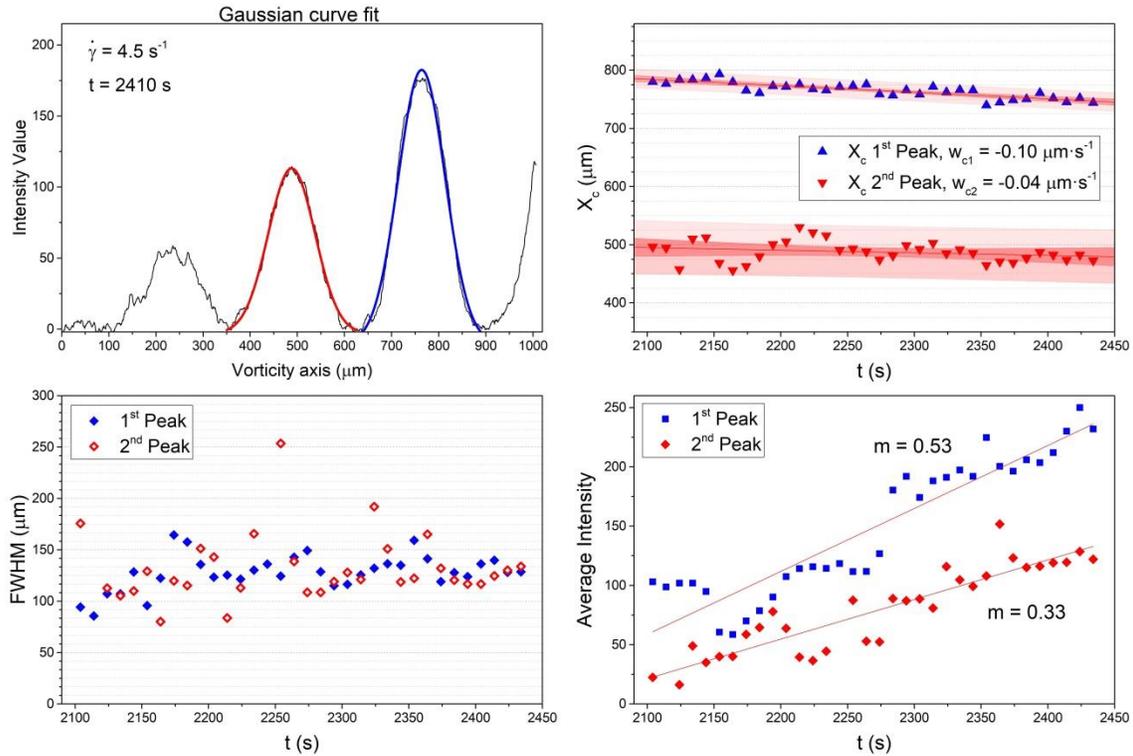

**Figure 9.** On the left upper side is the graph of intensity vs. vorticity axis (μm) of a pack of 10 images (1 fps) after a mean time of $t = 2410$ s from the beginning of the experiment, for to peaks. Top and right graph corresponds to the peak position displacement observed for all 34 ministacks, i.e., $x_c$ vs. time for the two most significant peaks. On the lower side are the graph of FWHM and of intensity vs. time, respectively. Blue data points correspond to the rightmost peak and red point to the middle peak.

For the analysis shown in Figure 9, the plots are the average intensity (vs. position along the vorticity axis) with an adjusted baseline. The *time tag* for this data set corresponds to 10 images, around 2410 s, after the onset of flow at $5\ \text{s}^{-1}$. These images were corresponding to 10 s of flow, with a total time of 337 images for the full stack. Thus, the information is coalesced into 34 mini-stacks (34 data points) to track the complete time evolution of the peak dynamics, i.e., duration of the experiment.

The blue and red solid traces (Figure 9a) are the adjusted Gaussian curves that best describes the intensities about those bands. Properties for the Gaussian trace indicate that its FWHM width remains roughly constant at 135 μm after 2150 s of flow, for both bands. In contrast, the growing amplitude indicates that the number of drops on the band is still increasing, but at a rate proportional to its height.

The trace of $x_{ci}$ vs. *time* measures the leftward displacement of the peak beginning at 2000 s after the onset of flow. The mean lateral displacement velocity of the peak is higher for the right peak at $-w_{c1} = 0.10$ μm·s$^{-1}$ (toward the center of rotation of the flow cell). In the Results and Discussion Sections, these findings are further analyzed.

## V.  RESULTS
### A. The Horizontal Distribution of Drops. The velocity field of the developed flow on the x-w plane

The observed accumulation of drops on a banded structure seems to imply that there is an underlying complex velocity field of the flow, possibly a fully 3-dimensional flow field. That is, on the one hand, for the simplest model for the viscosity of a mixture, $p \leq 1$ implies a lower viscosity for regions of a high fraction of aqueous phase. Thus, regions of low drop counts will indicate a rather high viscosity with respect to neighboring regions of high concentration of drops. On the other hand, if drop interactions are significant due to closeness among themselves, then a higher viscosity could be associated with a higher concentration of drops. In both cases, the relative viscosity along the vorticity axis should be an oscillating function of position. Hence the hor*izontal profile of the velocity field* along the flow direction may also show a sinusoidal variation along the z-axis—the vorticity axis—with a periodicity similar to that shown in Figure 8e.

Consequently, during the constant flow regime (towards the end of the transient state and for each shear rate), the velocity of individual drops along the flow direction is determined, and mean values calculated. As well, the spatial variability of the mean velocity —across columns that belong to valleys or peaks— is evaluated. For the present analysis, the time of each frame is 1 s.

| Shear Rate (s$^{-1}$) | Total Number Drops | Velocity Mean (μm/s) | Standard Deviation (μm/s) | Relative Deviation |
|---|---|---|---|---|
| 0.75 | 127 | 34.24 | 1.40 | 0.04 |
| 1.5 | 143 | 72.55 | 2.60 | 0.04 |
| 2.25 | 140 | 114.40 | 6.35 | 0.06 |
| 3 | 175 | 134.09 | 8.74 | 0.07 |
| 3.75 | 105 | 165.23 | 16.25 | 0.10 |
| 4.5 | 198 | 183.52 | 15.98 | 0.09 |

**Table 1.** The velocity of drops (randomly selected) observed for different shear rates.

Table 1 summarizes the observed velocities for a selection of drops for which its velocities are unambiguous. These velocity measurements are carried out with the ImageJ® software.

Figure 10a shows the mean velocity obtained for all shear rates, showing a linear behavior of the velocity with respect to $\dot{\gamma}$ for the weakest flows: $\dot{\gamma} \leq 2.25$ s$^{-1}$. The rate of increase slows down once the banded structure appears $\dot{\gamma} \geq 3.0$ s$^{-1}$, with twice the uncertainty of evaluated speeds (See Table 1; rightmost column).

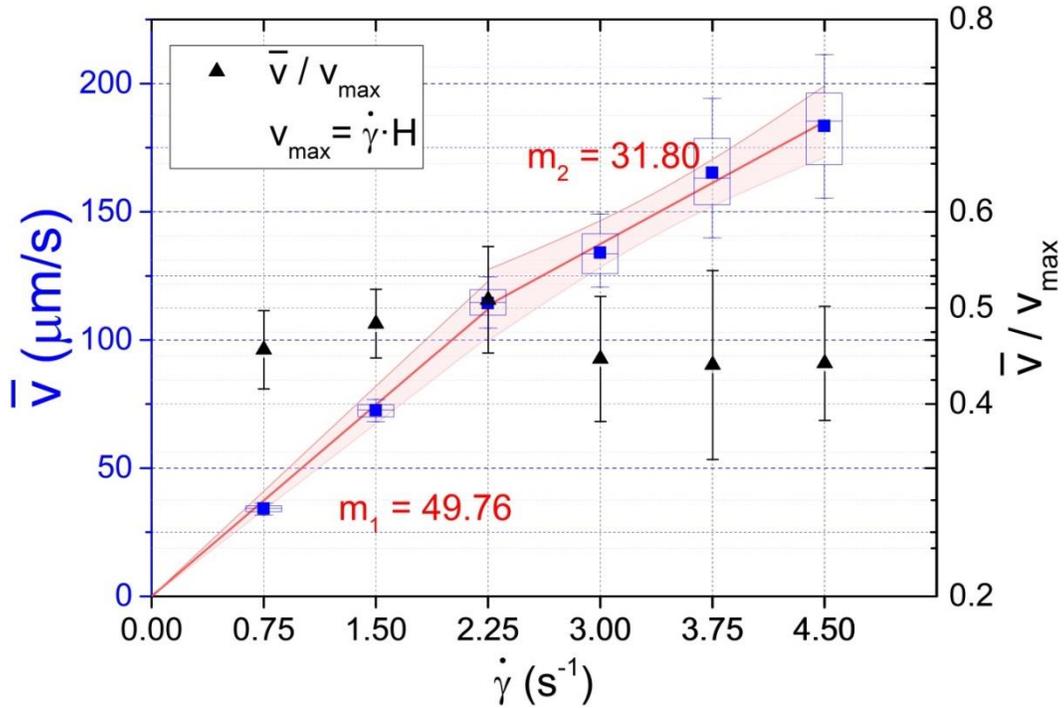

**Figure 10.** The average speed of drops over the complete flow domain, chosen for each shear rate: left ordinate. Also, the middle quartiles and the error bars are shown. The normalized dimensionless mean velocity wrt. the velocity of the shearing plate ($v_{max} = \dot{\gamma} \cdot H$); right ordinate. Slopes $m_1 = v \cdot \dot{\gamma}^{-1} = 50$ μm, and $m_2 = 31.8$ μm a test to a slow down of the flow speed once the banded structure of the emulsion occurs. Accordingly, the slow down of the normalized velocity is homogeneous and of about 10 % for the three highest shear rates.

The observed increase in the standard deviation values, as the shear rate increases, is due to intrinsic difficulties in evaluating the average speed of drops in a flow that is no longer laminar, as indicated in Table 1, and shown in Figure 9. Uncertainties for measured velocities increase, especially for those flow structures with a banded distribution of drops. This may imply two phenomena at play. *The first* one is simply an increase of collision rate between drops —in particular in the high concentration regime regions, inducing a slowdown of the measured velocity of individual drops. Moreover, *the second* may be due to a concentration-of-drops-dependent viscosity, with regions of low drop concentration associated with lower viscosity, thus, a higher velocity for a drop.

This scenario implies that the velocity profile for a lamella on the flow-vorticity plane shall show an oscillatory pattern as well. This pattern ought to be similar to the concentration of drops profile, normal to the vorticity axis. Even more, the velocity profile shall match the spacing of bands, and having higher velocities in regions of lower drop volumetric concentration.

In Figure 11, regions of low concentration do not have many available drops; hence, the number of data points is low. In contrast, regions of high concentration of drops may present many possible candidates for their velocity calculation, but constant interaction with many neighbors limits the number of useful candidates.

As well, Figure. 11 shows the oscillatory character of the *mean velocity of drops* within a given region, across the vorticity axis; see the red trace. Even more, the mean velocity profile shall match the spacing of bands and having higher velocities in regions of lower drop volumetric concentration, which can be inferred from the image of the banded structure of the emulsion at the bottom of the graph.

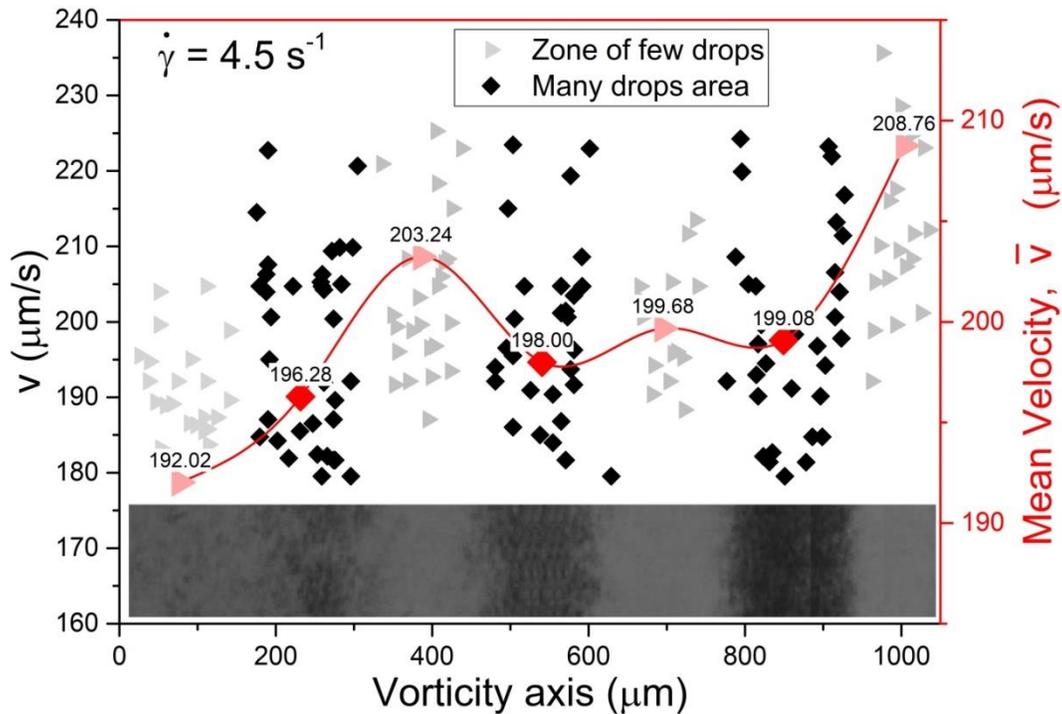

**Figure 11.** Drop velocities along the flow direction vs. their position across the vorticity axis. Left ordinate: plots the measured velocity for individual drops, for each marker. Right ordinate: the average velocity over the valley or peak of the drop distribution profile. Higher velocities coincide with regions of lower concentration of drops.

Thus, the information from Figures 8, 10 and 11 allow as to propose that the profile of velocities along the flow direction, spanning the full vorticity direction (that is, the *xw*-space or vorticity-flow plane), is non-homogenous, with a speed oscillation characterized by the same spatial frequency than the concentration of drops of the banded structure.

## B. The Vertical distribution of drops. The velocity field of the developed flow on the x-y plane

If the horizontal distribution of drops observed in Figure 8 induces a velocity distribution (an oscillatory velocity profile as shown in Figure 11) along the flow direction for all drops, then *the character of the flow field may be **fully three dimensional**,* even when it is generated by a flow cell with perfectly flat parallel surfaces. This assumption can be plausible due to several effects, which may occur simultaneously. Consequently, these phenomena may also indicate that *the vertical velocity profile is no longer linear*, mainly, due to a slowdown in the central lamella of the flow, as shown in Figure 10 for shear rates $\dot{\gamma} \geq 3.0 \text{ s}^{-1}$.

Even more, the slowdown of the mean flow is clearly non-homogeneous across the lamella, as seen in Figure. 11. Both pieces of information are very important and suggest that the velocity along the vorticity axis is no longer zero or homogeneous. In fact, this assumption could be plausible and could be explained by the oscillatory character of the flow field that induces a lateral component of the velocity —along the vorticity direction (inducing gradients of the concentration of drops, observed in the banded structure).

That is, there must be a component of the velocity field along the vorticity direction, which is weak but characterized by an oscillatory manner, as well. This normal component of the velocity field assures the development of a banded structure. The deceleration along the flow direction plus the appearance of a lateral (vorticity) component may also indicate that the vertical velocity profile will be three-dimensional.

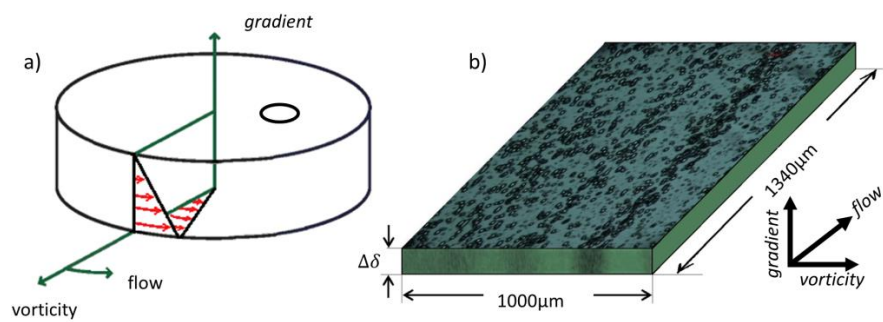

**Figure 12** (a) Scheme of the velocity profile in the flow-vorticity plane and in the gradient-flow plane, and (b) The captured image view in 3D: flow-vorticity plane about the center of the flow cell, thickness = $\Delta\delta$; and side, in the gradient-vorticity plane.

In order to study the profile of the concentration of drops and the vertical velocity profile of the bands, Figure 12 shows the geometry of the flow cell in three dimensions and *the initial assumption of a linear, unidirectional velocity profile*. In the lower part of the profile is the maximum speed and in the upper part the zero speed (static top plate). Figure 12 shows velocity measurements across the vorticity axis. Please note that the plate was translating at a constant velocity, i.e., the lower plate of the flow cell corresponds to the top value shown in Figure 13. It is possible to identify different bands that are separated by an approximate width of ~ 155 µm.

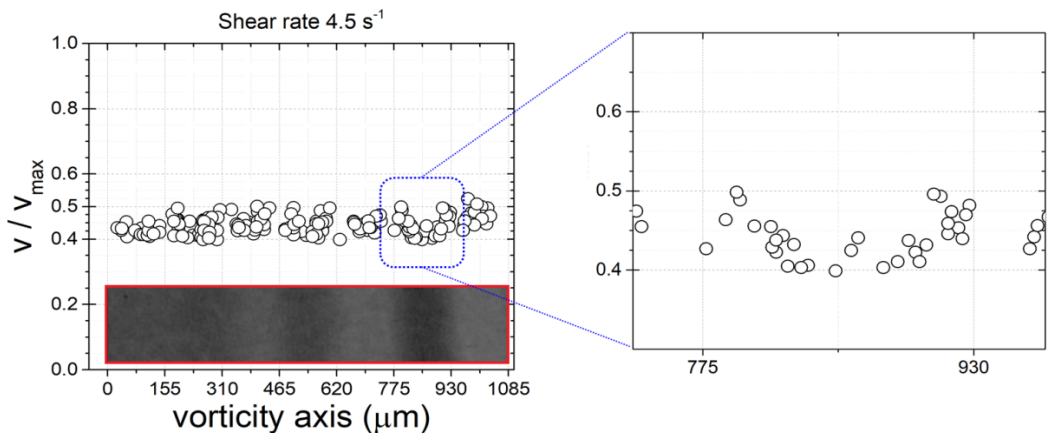

Figure 13 Dimensionless speed across the vorticity direction. Bottom insert in (a) shows the zones of the better-defined bands. (b) it is an expanded view of the selected region (blue square region) of the densest band.

Carrying out a characterization of *the vertical velocity of drops profile,* as shown in Figure 13, it is now possible to determine *the profile of the concentration of drops in a band, but now **in the gradient-vorticity plane**,* at least for the better-defined bands. That is, this analysis attempts to elucidate whether the band structure occurs from plate to plate of the shearing cell. However, given that the concentration of drops is quite large—actually, sufficiently high to preclude observation of the velocity of drops near the bottom of the flow cell—, then the measurement of velocities of drops corresponds to those in the upper half of the flow field, only.

Thus, based on the velocity of individual drops and using its normalized velocity to infer its vertical position on the band, it is possible to propose a vertical profile for the concentration of drops within a high concentration band. Figure 13. shows the possible position of drops and the corresponding upper-layer where drops are located.

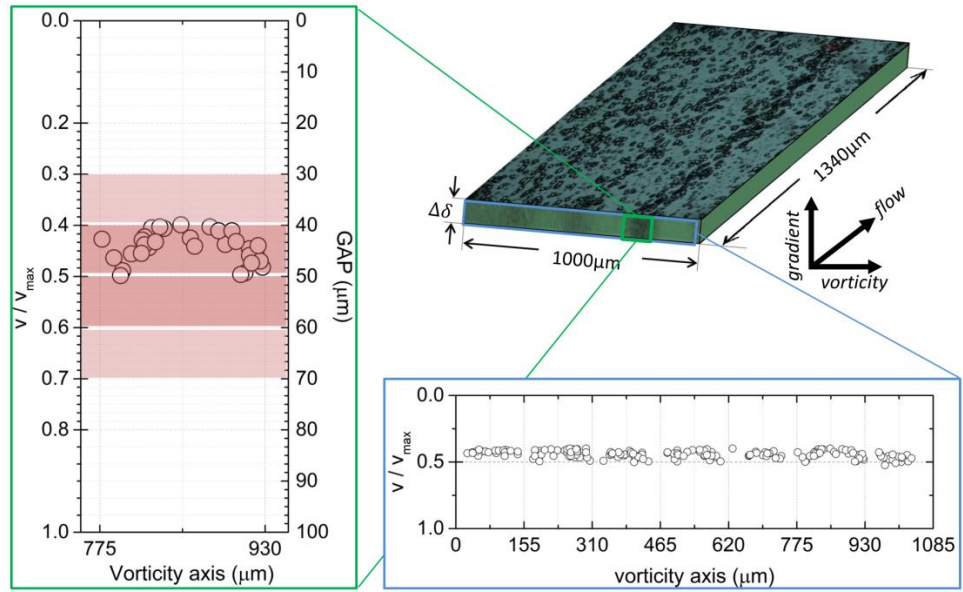

**Figure 14** Determination of the thickness of the band in the vorticity-gradient plane.

The concentration of drops inside the dotted red lamella, shown in Figure 14, is posited by symmetry considerations of what can only be determined with the measurement of the speed of drops. The red layer is the thickness of the band without taking into account the diameters of the drops. At the same time, the green one corresponds to the suspected thickness of the high concentration of drops layer when the diameter of the drops is taken into account. The observed diameter of droplets within this band area is of the order of 17 to 20 µm and is also taken into account to delimit the green layer thickness. In this way, a possible (actually the minimum) complete profile is predicated, which is only a fraction of the wall-to-wall separation, implying rather thick regions above and below with only a few drops (while the lamella maintains a high concentration of drops). The confinement parameter of the band as a whole is of Co = 0.4, which is considered in the literature as moderate confinement [21].

## VI. DISCUSSION OF RESULTS
### A. Critical capillary in concentrated emulsions

For a single drop embedded in a continuum, with a viscosity ratio of $p \sim 0.28$ and drop size $r \sim 18$ µm; the literature estimates that reasonable values of the critical capillary number in simple shear flow are about $Ca_{cr} \sim 0.51$ [22]. In the present work, the critical capillary is $Ca_{cr} \cong 0.21 \pm 0.07$, which may imply that other perturbation from nearby drops can induce rupture of drops at a lower $Ca_{cr}$.

Recalling Taylor's model prediction for the critical capillary, for a system with a constant value of $p$, the value of $Ca_{cr}$ implies that the critical radius (the largest radius value up to drops of stable shape) and the shear rate are inversely proportional. The observed discrepancy of these two values can now be used to understand a portion of the dynamics observed in relation to the bands structure.

In other words, the critical radius of a drop decreases as the shear rate increases, as shown in Figure 15 (the region delimited between the black dashed lines). These upper and lower limits of the critical diameter values are analytical results valid for slow flows, according to Taylor's predictions. This Figure also shows the complete evolution of the histograms of the drop size distribution for the full set of shear rates studied; the colored information portrays the histograms evaluated in Figures 3.4 and 3.5 and plotted vs. $\dot{\gamma}$.

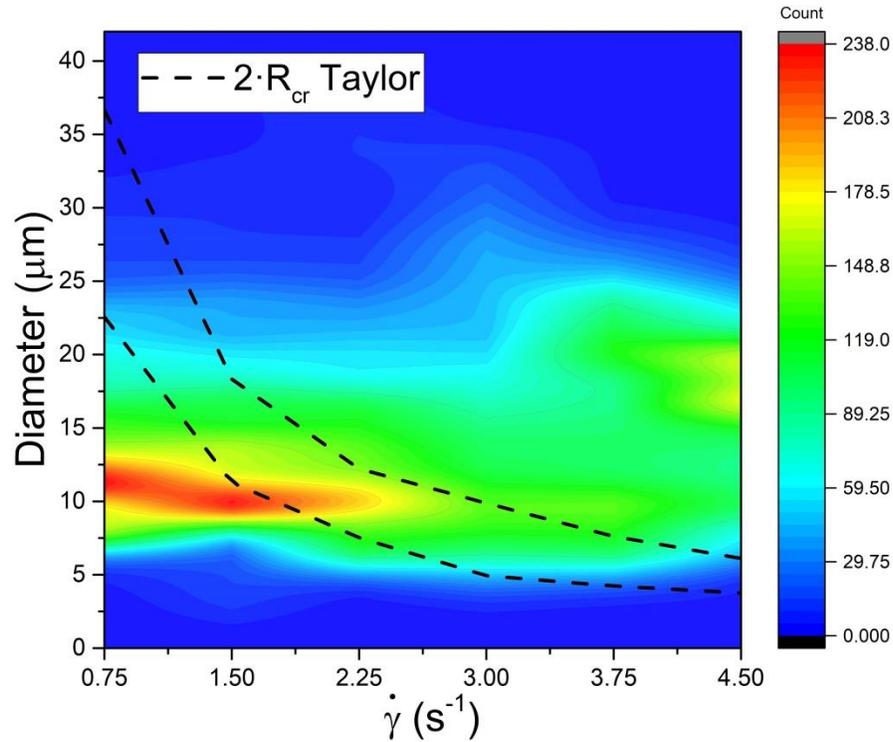

**Figure 15.** Distribution of size of drops in the emulsion vs. the complete range of shear rates used in the experiments; frequency of drops – color-coded. Comparison of the $Ca_{cr}$ obtained for Taylor's model: Black dashed region; the delimited area represents the range of stability for a single drop.

Please note that no drops smaller than 5 μm appear, as well as drops larger than 35 μm. The highest frequency occurs at low shear rates and for diameters of about 12 μm. Drop elongation and rupture of these drops does not occur until $\dot{\gamma} = 1.75 \text{ s}^{-1}$, when $Ca_{cr} \sim Ca_{Taylor}$. Higher shear rates preclude the observation of drops with a diameter of approx. 12 μm, hence a coalescence mechanism for growth of drops must be balanced by another of elongation and rupture of drops.

For shear rates $\dot{\gamma} \leq 2.75 \text{ s}^{-1}$, no drop shall be stable for diameters larger than 12 μm, implying that the observed stability of larger drops (i.e., about 16-20 μm) should be due to other stabilizing phenomena, mainly from nearby drops and a more complex flow regime. And for $\dot{\gamma} \sim 5 \text{ s}^{-1}$. Drops larger than 8 μm only exist when strong interaction with other drops occurs, and the flow regime is more complex than simple shear flow. Even more, these mechanisms appear to inhibit the existence of a drop larger than 25 μm.

Figure 16 shows the location of the more massive drops inside the image, which are mostly contained within the high concentration band, while smaller ones appear mainly inside the valleys between bands. The average diameter of the drops inside the bands is above the critical diameter of the single drop Taylor model (See Figure 15). In this way, $Ca_{cr} < Ca_{Taylor}$ even in this dilute regime, and $Ca_{cr}'' Ca_{Taylor}$ for drops confined in concentrated regions.

Therefore, in this paper is presented another possible explanation for the observed behavior in concentrated emulsions with bands present. This explanation is based more on the fact that the observed concentration gradients are concomitant of correlated gradients of the velocity field along the direction of the flow, as well as the appearance of a non-zero component of the velocity field along the direction of the vorticity.

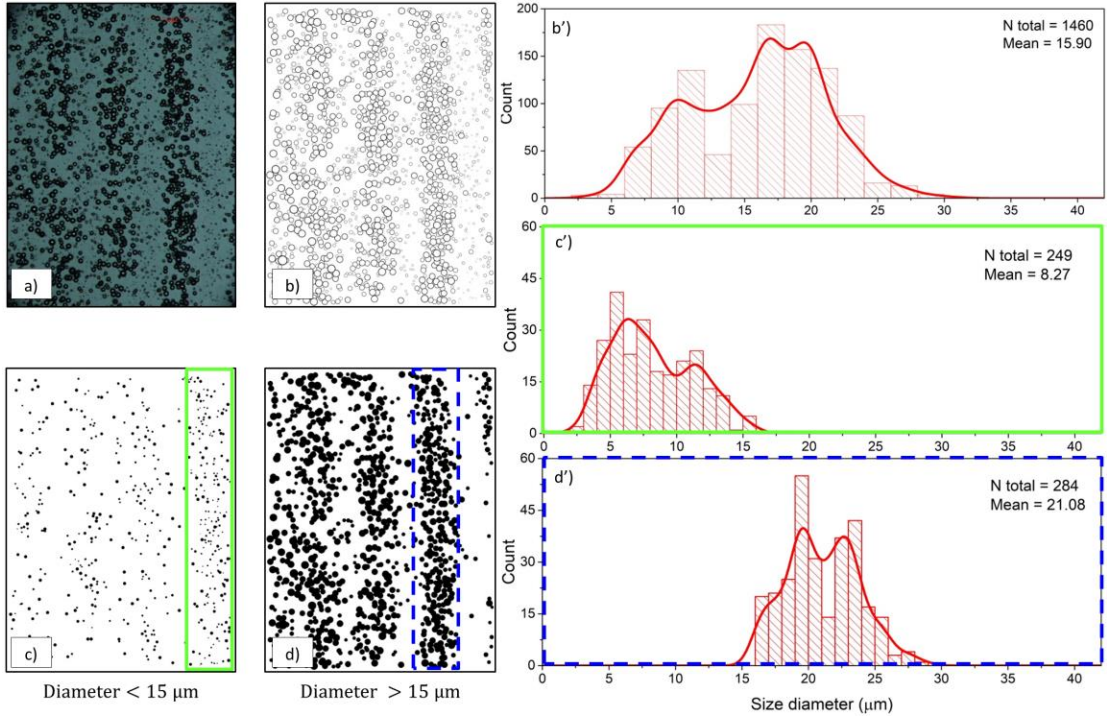

**Figure 16.** Correlation of position of a drop inside a slice of the flow-vorticity plane and the expected diameter of the drop. a) experimental data; b) processed image selecting spherical drops only: the sum of all-spherical drops, color-coded by size: total number of drops counted $n_{drops} = 1460$, with volume $V_{drops} = 6 \times 10^7$ μm$^3$ ; c) selection of drops with diameters < 15 μm, contained in the green area; d) selection of drops of diameters > 15 μm, contained in the blue area; and b') initial (top) histogram for drops after 2410 s, at $\dot{\gamma} = 5$ s$^{-1}$, c') histogram of drops in valleys (middle) and d'') histogram for flow region of high concentration (bottom).

Previous attempts to explain the observed distributions of drops were based upon the critical capillary number criteria. That is, this idea was reported by Sundararaj and Macosko in 1995 [23] and rests upon the assumption that emulsions are slightly concentrated systems, where the limiting case of Taylor's model can be referenced. Nevertheless, it is important to emphasize that the opposite is also stated in the literature by Jansen in 2001 [17]. In this work, drops inside the bands exceed the critical size due to a dynamic equilibrium, between the mechanisms that modify the morphology and the stress fields (by nearby neighbors, non-shearing motion), and eventually increases *the mean viscosity within the band* [24] [25].

The probability of contact for drops inside the bands is highest, giving coalescence a probable role, while close neighbors modified the stress field about a drop, decreasing rupture. It is then plausible that the average diameter of drops grows to a value that would not be expected, taking into account a reduced model for the emulsion, and more if we compare it with the most straightforward system: Taylor's model. However, a band of drops may impose a flow regime outside its core that is similar to that of a drop in the string of necklace form.

In Pathak's observations [26], four regimes are proposed, shown in Figure 17. Pathak attempts to predict possible morphologies in concentrated dispersions. The regimes classification is based on the dimensionless and *normalized* shear rate and the volumetric relation of the diameter of drops $D_{4,3}/H$. The first regime corresponds to the capillary number of drops exceeding the critical capillary, while the diameter of the drops is much less than the separation of the plates. The second regime corresponds to the case when the capillary number of drops is less than the critical capillary, and they are sufficiently dispersed that no hydrodynamic interaction is relevant in relation to neighbors or walls. The third regime is the case of highly deformed drops (i.e., oblate ellipsoids). The fourth regime is the case where the capillary number of drops (based on Taylor's model) is higher than the critical capillary, and the diameter of drops is greater than the separation of the plates. In this regime, drops with a highly deformed shape —strings or necklaces– are present, and provide a similar hydrodynamic environment to the banded structure of the emulsion. Moreover, the last regime (the fifth) is for the case where the diameter of the strings is below the separation of the plates, but the associated capillary number when strings are observed is less than the critical capillary of the strings, in conditions of low confinement.

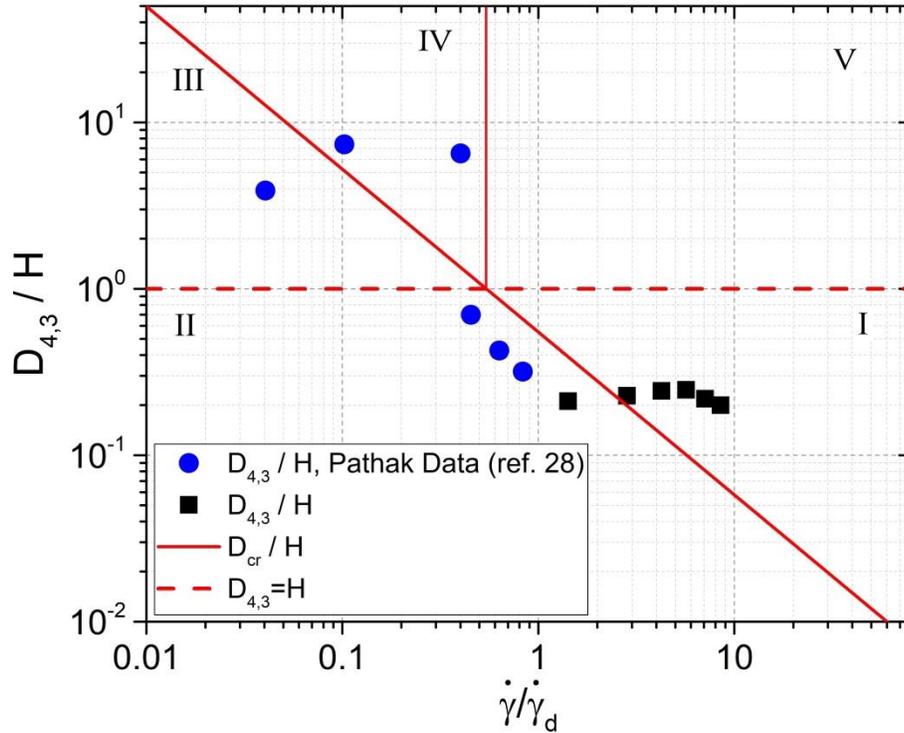

**Figure 17.** Adaptation of Pathak's representation of the regimes in the scenarios of concentrated emulsion morphologies. $\dot{\gamma}_d$ is the shear rate at which Taylor theory predicts the maximum droplet radius ($\dot{\gamma}_d = \sigma_{12}/(H \cdot \eta_m)$).

In Figure 17, the results of the experiments reported in this paper are indicated with data (black squares) points. Pathak's data [26] correspond to data (blue circles) points and corresponds to the behavior observed when an emulsion shows long strings. In principle, the results presented here correspond to Pathak's First Case, where drops are agglutinated and unstable because they stretch and break. However, the detailed structure and possible flow consideration that may induce such spatial distributions have not been studied experimentally, as far as it has been consulted in the literature, which is significant.

These data clearly do not match, but under these conditions, both strings, like bands, present stability, which might occur due to the influence of the confinement of the effect of the walls (on the band or the strings). This confinement leads to the existence of large drops that overcome the rupture, in both cases: Pathak with strings and the work presented in this paper with bands. These anisotropic structures are undoubtedly part of a transient state, which can be a consequence of deformations in bulk or in a concentrated emulsion.

The width of the bands is approximately 155 µm measured directly from the images. The bands move in a centripetal direction. The centripetal movement is a product of the accommodation of the areas of higher viscosity and those of lower viscosity [27] [28] [29]. At the present time, it is not possible to propose a mechanism that explains this phenomenon.

## VII. CONCLUSIONS

The formation of bands, perpendicular to the direction of the vorticity axis, is not well understood [3] [5] [6] [12] [29] [31]. The phenomenon of band formation is produced with relative simplicity, as is the case when mixing two immiscible Newtonian fluids, with a low viscosity ratio $p < 1$, without surfactants and for specific shear rate values [7]. A relatively plausible explanation is to attribute to the competing effect of coalescence and rupture, wherein values of $p$ less than one the coalescence predominate [16], and then the formation of large drops of a size determined by the separation of the discs [8], will end in the formation of the pearl neck structures, as evidenced in the images [19] However, it is an open subject of study.

In this paper, it is hypothesized that the curvature of the flow field has no significant role in the formation of bands, as posited by Caserta et al. [12] [29] —although more work is required and thus still unconfirmed— work that is projected into the future. Extrapolating the observations and taking as reference the work of Jeffrey Byars [28], it can be hypothesized that the observed bands correspond to a sector of an Archimedean spiral. We still cannot answer this question: if these bands are concentric rings or are spirals and this quest was not part of the objectives of this work.

However, here we show that a detailed analysis of the dynamics of the bands structures is possible, as well as the measurement of the flow field anomalies that are simultaneously observed. The *local viscosity* of the emulsion increases in areas of higher concentration of drops and regions of lower concentration of drops imply a lower viscosity, which causes the morphology of the emulsion to change macroscopically.

## VIII. ACKNOWLEDGMENT

E.L.M. thanks for financial support throughout his Ph.D. studies in Materials Science and Engineering, UNAM, to Consejo Nacional de Ciencia y Tecnología (CONACYT, Mexico). Also acknowledge finalcial support from DGAPA-UNAM, Mexico; grant number PAPIIT-IN114618.

21, p. 133, 2006 Oct 1.